# Nanofocusing, shadowing, and electron mean free path in the photoemission from aerosol droplets

Ruth Signorell[1,*], Maximilian Goldmann[1], Bruce L. Yoder[1], Andras Bodi[2], Egor Chasovskikh[1], Lukas Lang[1], David Luckhaus[1]

[1] Department of Chemistry and Applied Biosciences, Laboratory of Physical Chemistry, ETH Zürich, Vladimir-Prelog-Weg 2, CH-8093, Zürich, Switzerland

[2] Paul Scherrer Institute, CH-5232, Villigen, Switzerland

* To whom correspondence should be addressed. E-mail: rsignorell@ethz.ch

**Abstract:** Angle-resolved photoelectron spectroscopy of aerosol droplets is a promising method for the determination of electron mean free paths in liquids. It is particularly attractive for volatile liquids, such as water. Here we report the first angle-resolved photoelectron images of droplets with defined sizes, *viz.* of water, glycerol, and dioctyl phthalate droplets. Example simulations of water droplet photoelectron images and data for electron mean free paths for liquid water at low kinetic energy (< 3eV) are provided. We present an approach that allows one to gradually vary the conditions from shadowing to nanofocusing to optimize the information content contained in the photoelectron images

The emission of photoelectrons from small dielectric particles induced by ionizing radiation (referred to as "photoemission") is determined by the intensity distribution of the ionizing radiation and the electron migration and emission processes inside the particles. Variations in the internal light intensity and the electron migration processes modify the total photoelectron yield (PEY), the kinetic energy distribution of the photoelectrons (eKE), and their angular distribution (PAD). Figure 1 illustrates the influence of the variation of the light intensity inside a droplet (panels a and c) on velocity map photoelectron images (panels b and d) for dioctyl phthalate droplets of 250 nm radius for two limiting cases, which we refer to as "shadowing" (panels a and b) and "nanofocusing" (panels c and d). Electrons are preferentially formed in the regions inside a particle where the light intensity is high. Shadowing in the photoemission originates from strong light absorption on the side of the particle that faces the incoming light (Figure 1a). As a consequence, photoelectrons are preferentially emitted from this side of the particle and the highest photoelectron intensity is detected in the left half plane of the photoelectron image (Figure 1b). The opposite behavior, nanofocusing, occurs when the droplet acts as an optical resonator (Mie resonances; [1]). As a result intensity hot spots can form inside the droplet, for example in Figure 1c on the far side of the incoming light. Nanofocusing leads to preferential emission of photoelectrons in the propagation direction of the light; i. e. the highest electron intensity is recorded in the right half plane of the photoelectron image (Figure 1d). For very pronounced nanofocusing, the average light intensity inside the droplet is enhanced (typical enhancement factors of up to 10) compared with the intensity of the incident light, while shadowing leads to attenuation of the light inside the droplet compared with the intensity of the incoming light [1].

Watson was probably the first to study shadowing effects in solid dielectric particles in more detail in the context of vacuum ultraviolet (VUV) photoelectron emission from interstellar dust grains [2]. In the following years, photoemission from small particles was only sporadically addressed in various contexts. For example, enhanced photoemission from small aerosol particles was exploited by Burtscher, Siegmann, and Schmidt-Ott for aerosol analysis [3]. Ziemann and McMurry performed secondary electron yield measurements for probing organic films on aerosol particles [4] and recently Su *et al.* used photoelectron spectroscopy to study solvation phenomena in droplets [5]. Plasma formation in clusters and different types of particles after exposure to intense femtosecond pulses was observed through mass spectrometry and ion imaging [6,7], while electron imaging was used to study directional emission from dielectric particles using intense few-cycle laser fields [8,9]. Wilson et al. [10,11], the DESIRS beamline team at SOLEIL [12] and our group [13] used angle-resolved photoelectron spectroscopy and VUV radiation to visualize shadowing effects in the photoelectron angular distributions of small dielectric particles. The introduction of velocity map imaging for probing particle photoemission was crucial because it is currently the only method that allows one to obtain detailed information on the PEY, the eKE, and the PAD for small dielectric particles (~10 to 300 nm in radius). In principle, this opens up the possibility to exploit aerosol photoemission for the study of other intriguing phenomena. We have recently proposed angle-resolved photoelectron spectroscopy of aerosol particles as a new way to determine the mean free path of electrons (eMFP) in condensed matter [13]. Another example from our group is the size-dependence of photokinetics in small droplets, where nanofocusing and shadowing play an important role [14].

The attractiveness of shadowing and nanofocusing in aerosol photoemission for the study of other phenomena, such as electron scattering in condensed matter, relies on the ability to vary the penetration depth and intensity distribution of the light inside the aerosol particle. For the determination of eMFPs, for example, this allows one to generate electrons at various distances relative to the surface and thus to modify the electron escape conditions. The maximum information content can be extracted from photoelectron images when the conditions can be broadly varied from dominant shadowing to dominant nanofocusing. In principle, gradual variations between shadowing and nanofocusing can be achieved by varying the particle size or the wavelength of the incoming light. However, a gradual variation from dominant shadowing to dominant nanofocusing is challenging and has not yet been demonstrated. Another aspect that makes aerosol photoemission imaging attractive for the study of electron scattering processes is its applicability to a wide range of different compounds. This includes in particular liquids, such as water, for which eMFPs for slow electrons are not yet available [15-18]. The present paper reports on the first photoelectron images of liquid droplets - notably water droplets – and presents eMFP data for water at low kinetic energy (< 3eV). Additionally, it provides the first experimental demonstration of the full tunability between dominant shadowing and dominant nanofocusing.

Details of the photoelectron velocity map imaging (VMI) spectrometers used in this study have been provided in previous publications [13,19-21]. A sketch of the setup is shown in the SI, Figure S1. Droplets of DOP (dioctyl phthalate; Merck-Schuchardt), GLY (glycerol, Sigma-Aldrich), and water ($H_2O$, Cayman Europe) with broad size distributions were generated in nebulizers or atomizers and entered the photoelectron spectrometer through an aerodynamic inlet device, which also acted as a size selector. We estimate a geometrical standard deviation of the droplet size distribution after the selector of ~1.3 and a droplet density of ~$10^5$ cm$^{-3}$. The threshold ionization energies of $H_2O$, DOP, and GLY are ~133 nm [22], ~151 nm, ~141 nm, respectively. The droplets are ionized inside the VMI optics either with photons from a ns UV laser (tunable between 203 and 266 nm), or from a home-built tunable ns VUV laser [19], or from the VUV beamline of the Swiss Light Source at the Paul Scherrer Institute [23,24]. All data recorded with the ns UV laser correspond to two-photon ionization processes. All data recorded with the ns VUV laser and the continuous VUV beamline radiation are single-photon ionization processes. The VUV beamline has the advantage of easier tunability of the wavelengths while the VUV laser has the advantage that it is available in-house. The generated photoelectrons were projected onto two-dimensional (2D) electron imaging detectors. The distribution of the electron intensity in the photoelectron images is quantified by the asymmetry parameter $X$, defined as the difference between the fraction of electrons detected in the left and right half-plane of the photoelectron image.

$$X = \frac{I_l - I_r}{I_l + I_r},$$    Eq. (1)

where $I_l$ and $I_r$ are the integrated electron intensities in the left and right half plane, respectively, excluding the strong center spot. For details on the calculations of the internal electric field intensities (light intensity inside the droplet), the modelling of the photoelectron images for $H_2O$, and the extraction of eMFPs for $H_2O$ we refer the reader to the SI, Figure S3 and section B and ref. [13]. Briefly, the internal electromagnetic field intensity is calculated with classical electrodynamics using frequency-dependent refractive index data [1,25,26] and SI, Table T1. For single-photon ionization by the ns VUV laser and the synchrotron radiation, we display the internal field intensity while we show the square of the intensity for two-photon ionization by the ns UV laser. The migration of the electrons in the water droplets is described by Monte Carlo trajectories with elastic and inelastic cross sections describing the scattering events (SI, section B).

The degree of nanofocusing and shadowing in the light intensity (Figure 1a and c) is determined by the droplet radius $R$, the complex index of refraction of the droplet $N = n + ik$ and the wavelength of the light $\lambda$. The surroundings of the particle is vacuum. Shadowing dominates for absorption lengths $L_{\text{abs}} = \frac{\lambda}{4\pi k} \ll R$. This phenomenon is not limited to particles, but it is also observed in bulk. Nanofocusing, by contrast, is a resonance phenomenon in larger droplets, which does not occur in bulk or in very small droplets; e.g. in dipole scatterers [1]. It can only dominate when $\frac{\lambda}{2\pi|N|} \lesssim R \lesssim L_{\text{abs}}$; i. e. when resonances can build up ($\frac{\lambda}{2\pi|N|} \lesssim R$) and shadowing is of minor importance ($R \lesssim L_{\text{abs}}$) [1]. These general considerations imply that tuning between shadowing and nanofocusing should in principle be possible either by varying the particle size $R$ or the wavelength $\lambda$ of the light. However, as we discuss in the following, this depends on whether single-photon ionization in the VUV or two-photon ionization in the UV is used.

We first address the case of single-photon ionization in the VUV. An estimate of the above quantities for typical refractive indices at VUV wavelengths (e. g. refs. [25-27]) immediately reveals that nanofocusing does not usually occur for excitation with VUV light. For photon energies between the ionization energy and a few ten eV (wavelengths of a few ten nanometers), shadowing dominates for particle sizes above a few ten nanometers due to pronounced light absorption (high imaginary refractive index $k$; $L_{\text{abs}} = \frac{\lambda}{4\pi k} \ll R$). Droplets smaller than a few ten nanometers do not absorb strongly and thus do not show strong shadowing (see e.g. Fig. 1 in ref. [13]). However, these small droplets also do not exhibit nanofocusing because their size is below the limit where resonance phenomena occur ($\frac{\lambda}{2\pi|N|} > R$). At photon energies higher than a few ten eV, where $n < 1$ [26], nanofocusing is not observable either. As a result, we find that for many substances it is not possible to realize nanofocusing with single-photon VUV excitation, neither by variation of the wavelengths nor by variation of the particle size. Consequently, it is not possible to vary the conditions from shadowing to nanofocusing by VUV excitation. Figure 2 illustrates the wavelength-dependence of the calculated internal light intensity (left column) and the experimental photoelectron image (right column) for GLY droplets of 250 nm radius for five different

VUV wavelengths between 89 and 124 nm. This example reveals that both internal light intensity and photoelectron image are dominated by shadowing and that their spatial distribution depends only slightly on the wavelength. The wavelength-dependent refractive index $N = n + \mathrm{i}k$ and the asymmetry parameter $X$ (Eq. (1)) determined from the photoelectron images in the right column in Figure 2 are provided in Figures 3a and b, respectively. The asymmetry parameter in Figure 3b reveals that the information in the PADs (Figure 2, right row) is essentially identical for the different wavelengths. Note that the size of the electron images in Figure 2 decreases slightly with increasing wavelength simply because the maximum electron kinetic energy decreases for images recorded at lower photon energy. The behavior in Figures 2 and 3b is typical for single-photon VUV excitation. It is also found for many other compounds, such as KCl, water, benzene etc. [25-27], where $k$ values are high leading to predominant shadowing. Even though tuning from shadowing to nanofocusing is not feasible with VUV light, recent studies on NaCl and KCl particles demonstrate that it is still possible to vary the degree of shadowing from minor to pronounced shadowing by changing the particle size [10,13]. As demonstrated in ref. [13] for KCl particles and below for water droplets, this is already sufficient to retrieve eMFPs from particle photoelectron images.

As just explained, nanofocusing does not occur when VUV light is used. We thus report here a new approach that allows us to realize nanofocusing, and thus to gradually tune the conditions from shadowing to nanofocusing in a straightforward way. Instead of single-photon ionization by VUV light we use two-photon ionization by UV light. Many compounds have absorption bands in the UV, in the region of which the imaginary refractive index $k$ varies substantially within a narrow wavelength range of only a few nanometers. This is shown for the example of DOP in Figure 3c (see also SI, Table T1). At short wavelengths (e.g. 203 nm) where $k$ is high and absorption in the droplets is strong shadowing is pronounced, while at longer wavelength (e.g. 211 nm or 266 nm) light absorption is minor so that nanofocusing dominates. For wavelengths in between, there is a gradual change from dominant shadowing to dominant nanofocusing. This gradual change is illustrated in Figure 4 for DOP droplets of 250 nm radius for the calculated internal light intensity (left row) and the experimental photoelectron images (right row). With increasing wavelength (top to bottom), the center of gravity of the electron intensity (right row) shifts from the left half plane to the right half plane of the photoelectron image, in agreement with the shift of the intensity maximum of the light (left row) from the left side of the droplet to its right side. Figure 3d quantifies this gradual shift in the electron intensity in terms of the asymmetry parameter $X$, which in contrast to GLY droplets (Figure 3b) now changes pronouncedly from a positive to a negative value. To the best of our knowledge, this is the first time that a controlled gradual change between shadowing and nanofocusing is observed in photoelectron images of small particles. Figure S2 in the SI addresses another aspect, which is well known [1] but not obvious from Figure 4 and Figure 1c and d. It shows a map of the average of the square of the light intensity inside DOP droplets, demonstrating that the change from shadowing (e. g. 203 nm) to nanofocusing (266 nm) is accompanied by a change from attenuation to enhancement of the average light intensity inside the droplet compared with the intensity of the incoming light. Note that the internal intensity enhancement inside the droplet for also results in an increase in the PEY; a phenomenon that does not occur in the bulk.

By choosing the proper UV wavelength, i.e. using a wavelength at which absorption is not too strong, it is in principle also possible to tune between shadowing and nanofocusing by changing the droplet size instead of the wavelengths as in Figure 4. However, to achieve full tunability from dominant shadowing to dominant nanofocusing a very broad size range needs to be covered. This is experimentally challenging because it requires methods for particle generation and size selection over a broad size range at high particle number concentrations.

For solid and non-volatile liquids atomizers and nebulizers are typically used in combination with scanning mobility particle sizers for size selection. The limitation for size variation of solid particles is usually the solubility in the solvent that is used for atomization/nebulization. For liquids, it is imperative that the proper aerodynamic inlet device is used and operated under conditions specified by its design. Otherwise, loss and coagulation of droplets can severely distort the droplet size and lower the droplet number density. Volatile liquids, such as water, come with the additional issue of evaporation in and after the aerodynamic device because of reduced pressure conditions. For all of these reasons, tuning the wavelengths in the UV remains the simplest way to gradually vary between dominant shadowing and dominant nanofocusing; provided of course that the compound of interest has appropriate UV absorption bands.

Given the above-mentioned experimental challenges associated with volatile liquids, it is not surprising that photoelectron imaging of water droplets has not yet been reported. Particularly challenging are quantitative studies on water droplets, such as the one reported below. They require narrow droplet size distributions of well-defined size and accurate information on the PEY, the eKE, and the PAD. The example for water droplets of 100 nm radius in Figure 5a and b demonstrates that photoelectron images of good quality can be recorded and that quantitative information – here on eMFPs – can be extracted from these droplet photoelectron images. Both the experimental and the calculated photoelectron image in panel a and b, respectively, show slight shadowing in agreement with the calculated internal light intensity displayed in panel c. The experimental photoelectron image agrees well with the calculated image. Histograms of the corresponding experimental and calculated 2D velocity distribution are provided in the SI in Figure S3.

The calculation of the electron image in Figure 5b requires modeling of the electron migration inside the droplets, which is determined by the eMFPs. We have recently demonstrated for KCl aerosol particles that eMFPs can be extracted from fits of calculated photoelectron images to experimental ones [13]. The photoelectron images in the present work show a strongly improved image quality compared with that proof-of-principle study. This was achieved by replacing the photoelectron spectrometer (imaging optics, detector) at the VUV beamline. In addition, we have replaced the rather simple model used in the proof-of-principle study by a detailed model for electron transport that includes the explicit treatment of the various scattering processes and their angular dependences. More information is provided in the SI in Section B (see also refs. [15,28]). The inelastic and *isotropic* elastic mean free paths and the electron attenuation lengths retrieved for liquid water at electron kinetic energies below 3 eV are listed in Table 1. (Note that the *anisotropic* elastic mean free path is dominated by forward scattering, which therefore does not affect the distribution of the photoelectrons [15,28].) These are the first eMFP data for liquid water at such low energies so that there are no previous data to compare with directly. Lübcke and coworkers recently reported a single value for the electron attenuation length (EAL) in liquid water of ~5 nm at low eKE from a liquid water microjet study [29]. This seems to be a very approximate value as the authors emphasize that their "experimental results are also consistent with a significantly larger EAL" without further specifying any uncertainties. The EAL can be calculated from the eMFP but not vice versa. Our model predicts EALs that are only slightly smaller than the inelastic eMFPs (see Table 1), which agrees well with the approximate EAL value from ref. [29]. Our results suggest a surprisingly robust scaling of the energy dependence of relative scattering cross section. Thus we find that the purely vibrational scattering (i.e. off intramolecular vibrational modes) increases by a factor of 1.2 when the kinetic energy of the electron decreases from 1.7 eV to 1.0 eV. In the gas phase the corresponding ratio amounts to 1.36 [Yukikazu Itikawa and Nigel Mason, J. Phys. Chem. Ref. Data, **34**, No.1, 1-22 (2005)]. In the same energy range, the ratio of (isotropic) elastic to pure vibrational scattering of electrons increases 2.4-fold in water vapour and 2.1-fold in the liquid. Moreover, the good agreement between our simulations and VMI experiments lend further support to the proposed similarity between the electron transport properties in liquid water and amorphous water *ice* [15].

Electron mean free path data at low electron kinetic energies are essential parameters needed for the description of the behavior of solvated electrons as well as for the quantitative modeling of radiation damage to biological tissues [29-38]. In this context, the data in Table 1 resolve long lasting discussions about the order of magnitude of the eMFP values at very low electron kinetic energies. We would also like to note that recently progress has been made in the determination of EALs at higher electron energies than the ones considered here using liquid water microjets [16-18]. We are currently extending our droplet studies to higher electron energies with the goal to reach a similar range of electron kinetic energies. This would allow one to compare eMPFs retrieved from two complementary approaches, which both have their advantages and disadvantages. A potential advantage of the aerosol approach over liquid jet studies might derive from the size-dependent information on the eKE and the PAD [13].

In summary, we present the first photoemission images of water and other aerosol droplets with defined droplet sizes, and demonstrate for the example of water droplets that low energy electron mean free paths can be obtained from a comparison of calculated and experimental images. The work shows that photoemission imaging of aerosol droplets has become a quantitative technique, which for volatile liquids provides a complementary approach to liquid microjet studies. Full tunability from dominant shadowing to dominant nanofocusing is demonstrated for the first time by means of two-photon ionization in the UV instead of single-photon VUV ionization. This straightforward approach allows one to maximize the information content that can be extracted from photoelectron imaging for compounds with UV absorption bands. More generally, photoelectron imaging and spectroscopy of droplets opens up a new avenue for the study of intriguing phenomena that range from electron migration processes and confinement effects in droplet aerosol kinetics to solvation phenomena and nanoplasma formation [5,7,14].

**Acknowledgment:** We are very grateful to David Stapfer and Markus Steger from our workshops for technical support and we thank Guido Grassi from our analytical services for assistance with the determination of the DOP refractive index data. Financial support was provided by the Swiss National Science Foundation under project no. 200020_159205, ETH Zurich, and the ETH-FAST initiative.

**LIST OF FIGURES**

**Table 1.**   Inelastic and *isotropic* elastic mean free path and electron attenuation lengths for liquid water retrieved from photoelectron imaging of water droplets. We estimate uncertainties of ~40% for the inelastic mean free path and a factor of two for the isotropic elastic mean free path.

| electron kinetic energy / eV | inelastic mean free path / nm | *isotropic* component of the elastic mean free path / nm | Electron attenuation length / nm |
|---|---|---|---|
| 3.0 | 5.1 | 16.2 | 3.9 |
| 2.5 | 4.5 | 15.2 | 3.5 |
| 2.0 | 3.9 | 14.8 | 3.1 |
| 1.5 | 3.4 | 11.1 | 2.6 |
| 1.0 | 2.9 | 5.3 | 2.2 |

# Figures

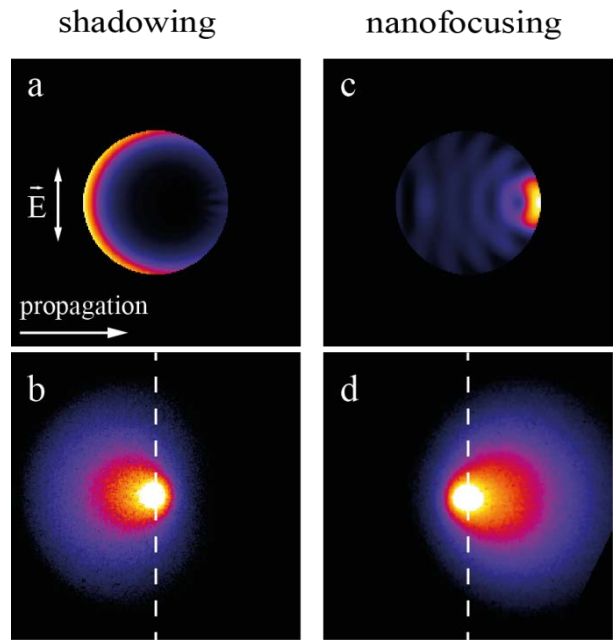

**Figure 1.** Dominant shadowing (a,b) and dominant nanofocusing (c,d) in dioctyl phthalate (DOP) droplets of 250 nm radius. (a,b) For single-photon excitation at a wavelength of 93 nm recorded with the ns VUV laser. (c,d) For two-photon excitation using 266 nm light from the ns UV laser. (a,c) Calculated light intensity (c: squared) inside the droplet. Shown is the intensity distribution of the ionizing radiation in a plane through the center of the droplet that is spanned by the polarization ($\vec{E}$) and propagation direction of the light. The polarization and propagation direction of the light are indicated by arrows in panel a. The light intensity decreases from yellow to red to blue. (b,d) Experimental photoelectron images with asymmetry parameters $X = 0.43$ and -0.47, respectively (see Eq. (1)). The dashed white line divides the left and half plane of the photoelectron image. The electron intensity decreases from yellow to red to blue.

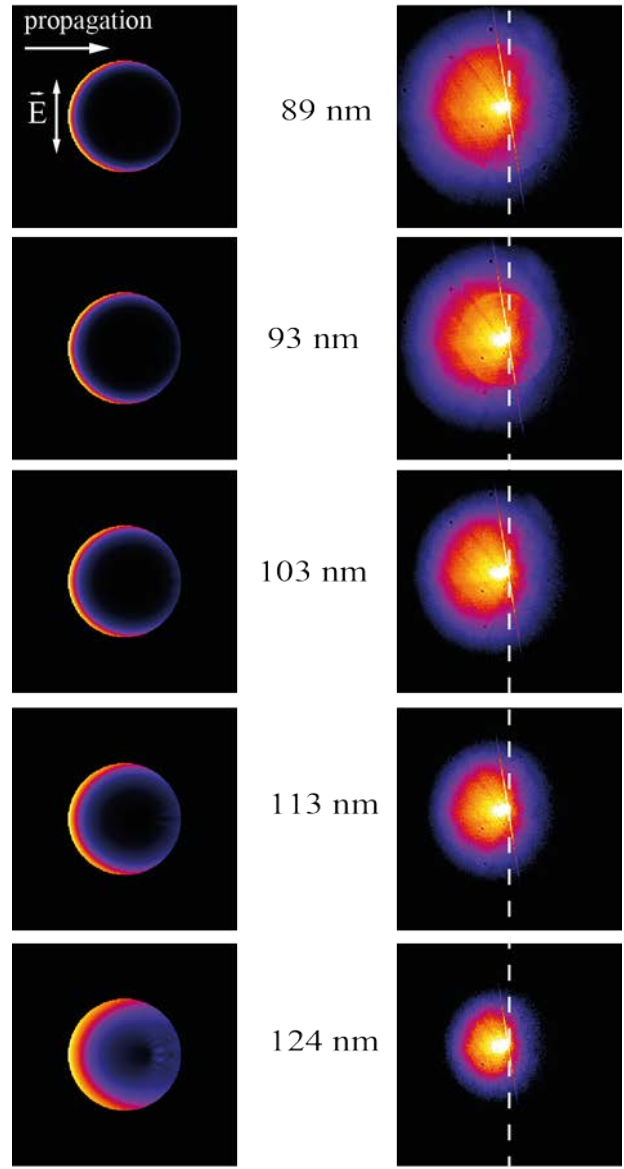

**Figure 2.** Dominant shadowing in glycerol (GLY) droplets of 250 nm radius for single-photon excitation at five different wavelengths between 89 and 124 nm (top to bottom) recorded at the VUV beamline. Left column: Calculated light intensities inside the droplet. The intensity distribution is shown in a plane through the center of the droplet that is spanned by the polarization ($\vec{E}$) and propagation direction of the light. Refractive index data are taken from ref.[25]. Right column: Experimental photoelectron images from single-photon ionization. The oblique lines that cross the images are an artefact of the delay-line detector. This artefact has a width of only a single pixel and can be removed by interpolation. The distance from the center to edge in the photoelectron images corresponds to an electron kinetic energy of 4 eV.

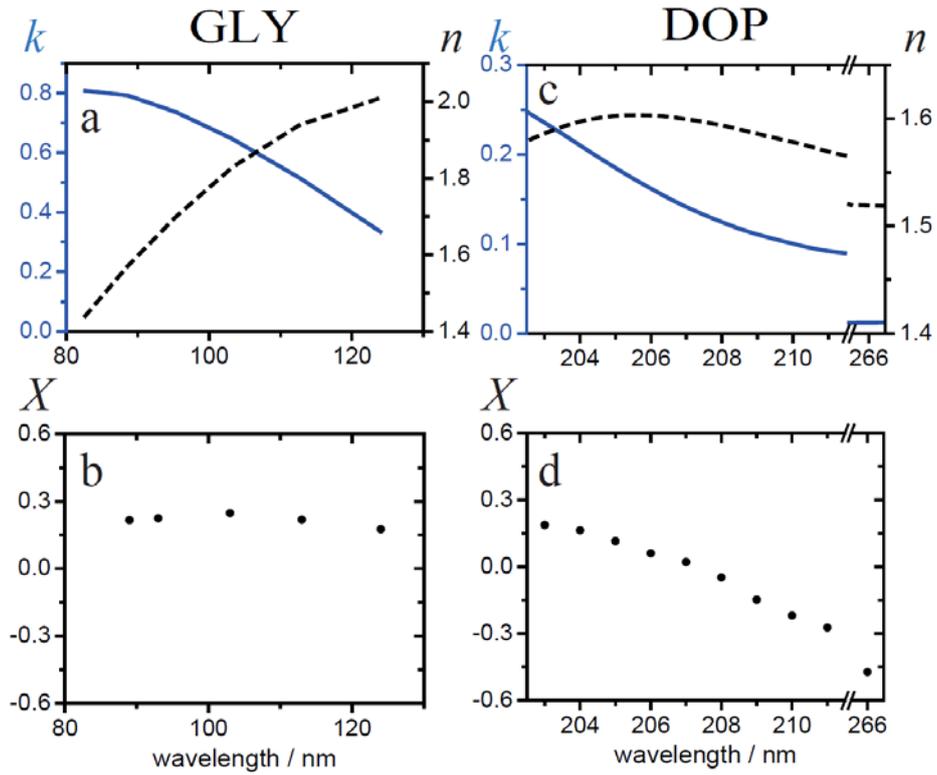

**Figure 3.** (a) Refractive index of GLY between 89 and 124 nm [25]. *n* (dashed black line) and *k* (full blue line) are the real and the imaginary part of the refractive index, respectively. (b) Asymmetry parameter *X* for the GLY photoelectron images in the right column of Figure 2. (c) Refractive index of DOP between 203 and 266 nm (see SI, Table T1) (d) Asymmetry parameter *X* for the DOP photoelectron images in the right column of Figure 4.

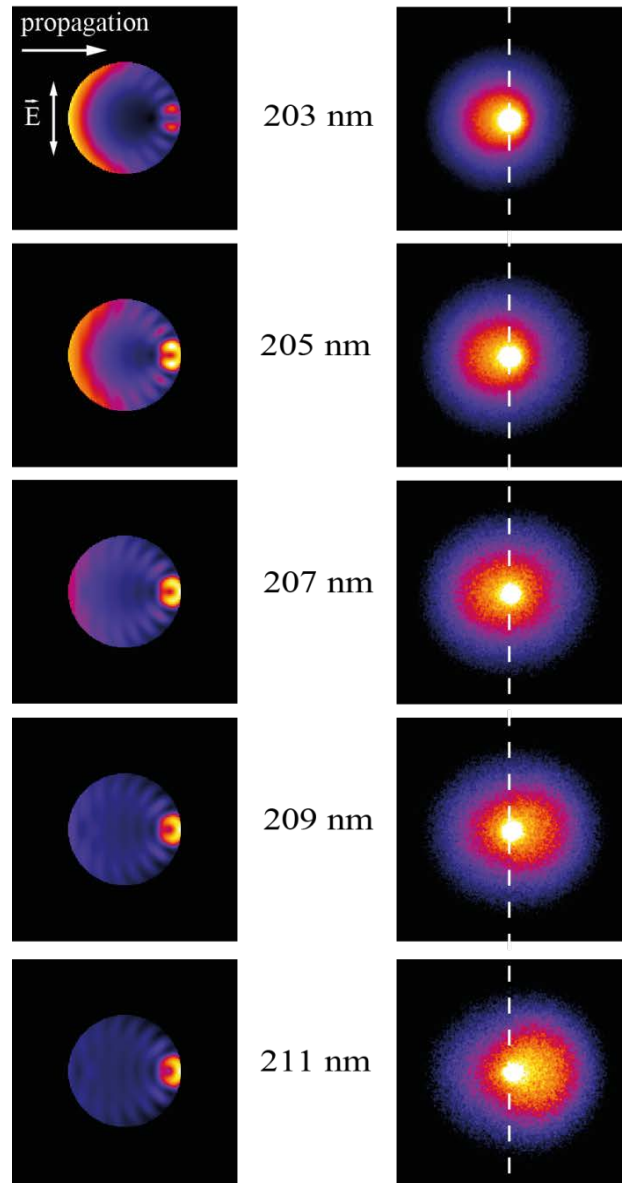

**Figure 4.** Gradual change from dominant shadowing to dominant nanofocusing for dioctyl phthalate droplets of 250 nm radius as a function of increasing wavelength for five different wavelengths between 203 and 211 nm (top to bottom) from the ns UV laser. Left column: Calculated light intensity (square) inside the droplet. The intensity distribution is shown in a plane through the center of the droplet that is spanned by the polarization and propagation direction of the light. Right column: Experimental photoelectron images. The droplets are ionized by two-photon processes. The distance from the center to edge in the photoelectron images corresponds to an electron kinetic energy of 0.9 eV.

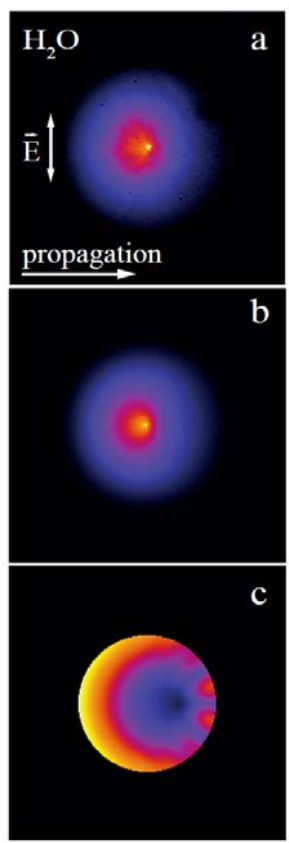

**Figure 5.** Shadowing in water droplets of 100 nm radius for single-photon excitation with VUV light of 103.3 nm. (a) Experimental photoelectron image for single-photon ionization. (b) Corresponding calculated photoelectron image for single-photon ionization (see SI, Figure S3 and Section B). This photoelectron spectrum was calculated with the eMFP data listed in Table 1. The total calculated electron yield is 20 %. The distance from the center to edge in the photoelectron images in panels a and b corresponds to an electron kinetic energy of 7.5 eV. (c) Calculated light intensity inside the droplet. The intensity distribution is shown in a plane through the center of the droplet that is spanned by the polarization and propagation direction of the light.

**REFERENCES**

[1] C.F. Bohren, D.R. Huffman, Absorption and Scattering of Light by Small Particles, John Wiley & Sons, New York, 1983.
[2] W.D. Watson, J. Opt. Soc. Am. 63 (1973) 164.
[3] H. Burtscher, L. Scherrer, H.C. Siegmann, A. Schmidt-Ott, B. Federer, J. Appl. Phys. 53 (1982) 3787.
[4] P.J. Ziemann, P.H. McMurry, Aerosol Sci. Technol. 28 (1998) 77.
[5] C.-C. Su, Y. Yu, P.-C. Chang, Y.-W. Chen, I.-Y. Chen, Y.-Y. Lee, C.C. Wang, J. Phys. Chem. Lett. 6 (2015) 817.
[6] S. Schorb, D. Rupp, M.L. Swiggers, R.N. Coffee, M. Messerschmidt, G. Williams, J.D. Bozek, S.-I. Wada, O. Kornilov, T. Möller, C. Bostedt, Phys. Rev. Lett. 108 (2012) 233401.
[7] D.D. Hickstein, F. Dollar, J.L. Ellis, K.J. Schnitzenbaumer, K.E. Keister, G.M. Petrov, C. Ding, B.B. Palm, J.A. Gaffney, M.E. Foord, S.B. Libby, G. Dukovic, J.L. Jimenez, H.C. Kapteyn, M.M. Murnane, W. Xiong, ACS Nano 8 (2014) 8810.
[8] S. Zherebtsov, T. Fennel, J. Plenge, E. Antonsson, I. Znakovskaya, A. Wirth, O. Herrwerth, F. Süssmann, C. Peltz, I. Ahmad, S.A. Trushin, V. Pervak, S. Karsch, M.J.J. Vrakking, B. Langer, C. Graf, M.I. Stockman, F. Krausz, E. Rühl, M.F. Kling, Nature Phys. 7 (2011) 656.
[9] F. Süssmann, L. Seiffert, S. Zherebtsov, V. Mondes, J. Stierle, M. Arbeiter, J. Plenge, P. Rupp, C. Peltz, A. Kessel, S.A. Trushin, B. Ahn, D. Kim, C. Graf, E. Rühl, M.F. Kling, T. Fennel, Nature Comm. 6 (2015) 7944.
[10] K.R. Wilson, S. Zou, J. Shu, E. Rühl, S.R. Leone, G.C. Schatz, M. Ahmed, Nano Lett. 7 (2007) 2014.
[11] M.J. Berg, K.R. Wilson, C.M. Sorensen, A. Chakrabarti, M. Ahmed, J. Quant. Spectrosc. Radiat. Transfer 113 (2012) 259.
[12] F. Gaie-Levrel, G.A. Garcia, M. Schwell, L. Nahon, Phys. Chem. Chem. Phys. 13 (2011) 7024.
[13] M. Goldmann, J. Miguel-Sánchez, A.H.C. West, B.L. Yoder, R. Signorell, J. Chem. Phys. 142 (2015) 224304.
[14] J.W. Cremer, K.M. Thaler, C. Haisch, R. Signorell, Nature Comm. 7 (2016) 10941.
[15] M. Michaud, A. Wen, L. Sanche, Radiat. Res. 159 (2003) 3.
[16] N. Ottosson, M. Faubel, S.E. Bradforth, P. Jungwirth, B. Winter, J. Electron Spectrosc. Relat. Phenom. 177 (2010) 60.
[17] S. Thürmer, R. Seidel, M. Faubel, W. Eberhardt, J.C. Hemminger, S.E. Bradforth, B. Winter, Phys. Rev. Lett. 111 (2013) 173005.
[18] Y.-I. Suzuki, K. Nishizawa, N. Kurahashi, T. Suzuki, Phys. Rev. E 90 (2014) 010302.
[19] B.L. Yoder, A.H.C. West, B. Schläppi, E. Chasovskikh, R. Signorell, J. Chem. Phys. 138 (2013) 044202.
[20] R. Signorell, B.L. Yoder, A.H.C. West, J.J. Ferreiro, C.-M. Saak, Chem. Sci. 5 (2014) 1283.
[21] A.H.C. West, B.L. Yoder, R. Signorell, J. Phys. Chem. A 117 (2013) 13326.
[22] B. Winter, R. Weber, W. Widdra, M. Dittmar, M. Faubel, I.V. Hertel, J. Phys. Chem. A 108 (2004) 2625.
[23] M. Johnson, A. Bodi, L. Schulz, T. Gerber, Nucl. Instrum. Methods Phys. Res., Sect. A 610 (2009) 597.
[24] A. Bodi, P. Hemberger, T. Gerber, B. Sztáray, Rev. Sci. Instrum. 83 (2012) 083105.
[25] R.D. Birkhoff, L.R. Painter, J.M. Heller, J. Chem. Phys. 69 (1978) 4185.
[26] H. Hayashi, N. Hiraoka, J. Phys. Chem. B 119 (2015) 5609.

[27]   D.M. Roessler, W.C. Walker, J. Opt. Soc. Am. 58 (1968) 279.
[28]   I. Plante, F.A. Cucinotta, New J. Phys. 11 (2009) 063047.
[29]   F. Buchner, T. Schultz, A. Lübcke, Phys. Chem. Chem. Phys. 14 (2012) 5837.
[30]   E. Alizadeh, L. Sanche, Chem. Rev. 112 (2012) 5578.
[31]   E. Alizadeh, T.M. Orlando, L. Sanche, Annu. Rev. Phys. Chem. 66 (2015) 379.
[32]   M. Faubel, K.R. Siefermann, Y. Liu, B. Abel, Acc. Chem. Res. 45 (2012) 120.
[33]   M.H. Elkins, H.L. Williams, A.T. Shreve, D.M. Neumark, Science 342 (2013) 1496.
[34]   Y.-I. Yamamoto, Y.-I. Suzuki, G. Tomasello, T. Horio, S. Karashima, R. Mitrić, T. Suzuki, Phys. Rev. Lett. 112 (2014) 187603.
[35]   A.H.C. West, B.L. Yoder, D. Luckhaus, C.-M. Saak, M. Doppelbauer, R. Signorell, J. Phys. Chem. Lett. 6 (2015) 1487.
[36]   R. Seidel, B. Winter, S. Bradforth, Annu. Rev. Phys. Chem. 67 (2016).
[37]   D.M. Sagar, C.D. Bain, J.R.R. Verlet, J. Am. Chem. Soc. 132 (2010) 6917.
[38]   J.M. Herbert, L.D. Jacobson, J. Phys. Chem. A 115 (2011) 14470.

**Supplementary Information for**

**Nanofocusing, shadowing, and electron mean free path in the photoemission from aerosol droplets**

Ruth Signorell[1,*], Maximilian Goldmann[1], Bruce L. Yoder[1], Andras Bodi[2], Egor Chasovskikh[1], Lukas Lang[1], David Luckhaus[1]

[1] *Department of Chemistry and Applied Biosciences, Laboratory of Physical Chemistry, ETH Zürich, Vladimir-Prelog-Weg 2, CH-8093, Zürich, Switzerland*

[2] *Paul Scherrer Institute, CH-5232, Villigen, Switzerland*

\* To whom correspondence should be addressed. E-mail: rsignorell@ethz.ch

## A. Additional Figures and Tables

**Figure S1: Sketch of the experimental setup.** Aerosol particles are generated either with an atomizer or nebulizer and introduced into vacuum via an aerodynamic inlet, which also acts as a size selector. The resulting beam of aerosol particles passes from the source chamber (labelled '1') to the detection chamber (labelled '2') before being photoionized by either UV or VUV light (directed perpendicular to the plane of the page, interaction volume labelled '↔'. The resulting photoelectrons are accelerated by a three plate extractor (labelled 'VMI optics') toward a position sensitive detector in a perpendicular arrangement.

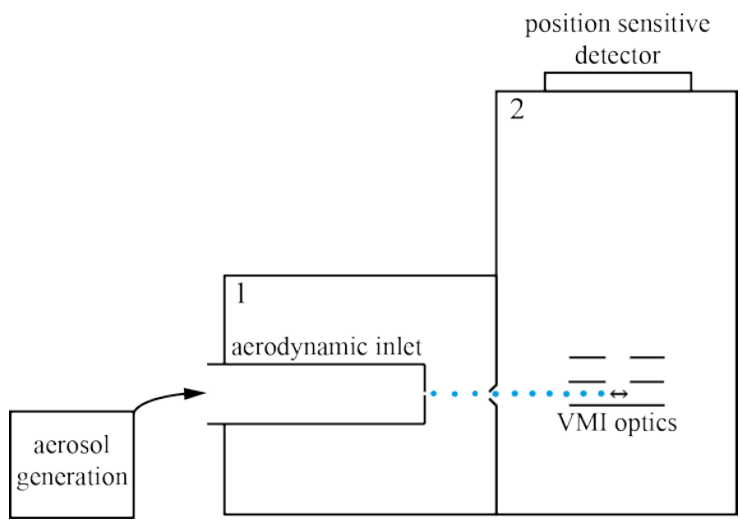

**Table T1: Refractive index data of DOP.** The real part $n$ and imaginary part $k$ of the complex index of refraction $N = n + ik$ were determined from UV/VIS absorption spectra of different DOP bulk solutions using Kramers-Kronig inversion [Hawranek *et al.*, Spectrochim. Acta A, 32, 85, (1976)].

| λ [nm] | n | k | λ [nm] | n | k |
|---|---|---|---|---|---|
| 800 | 1.49 | 0.00 | 231 | 1.56 | 0.08 |
| 660 | 1.49 | 0.00 | 230 | 1.56 | 0.08 |
| 580 | 1.49 | 0.00 | 229 | 1.55 | 0.08 |
| 500 | 1.49 | 0.00 | 228 | 1.55 | 0.08 |
| 420 | 1.49 | 0.00 | 227 | 1.55 | 0.09 |
| 380 | 1.49 | 0.00 | 226 | 1.54 | 0.09 |
| 360 | 1.50 | 0.00 | 225 | 1.54 | 0.09 |
| 340 | 1.50 | 0.00 | 224 | 1.54 | 0.09 |
| 320 | 1.50 | 0.00 | 223 | 1.53 | 0.09 |
| 300 | 1.51 | 0.00 | 222 | 1.53 | 0.09 |
| 290 | 1.52 | 0.00 | 221 | 1.53 | 0.09 |
| 280 | 1.52 | 0.01 | 220 | 1.53 | 0.09 |
| 275 | 1.52 | 0.02 | 219 | 1.53 | 0.08 |
| 270 | 1.52 | 0.01 | 218 | 1.53 | 0.08 |
| 266 | 1.52 | 0.01 | 217 | 1.53 | 0.08 |
| 262 | 1.52 | 0.01 | 216 | 1.54 | 0.08 |
| 260 | 1.53 | 0.01 | 215 | 1.54 | 0.08 |
| 258 | 1.53 | 0.01 | 214 | 1.55 | 0.08 |
| 256 | 1.53 | 0.01 | 213 | 1.56 | 0.08 |
| 254 | 1.53 | 0.01 | 212 | 1.56 | 0.09 |
| 252 | 1.54 | 0.01 | 211 | 1.57 | 0.09 |
| 250 | 1.54 | 0.02 | 210 | 1.58 | 0.10 |
| 248 | 1.54 | 0.02 | 209 | 1.59 | 0.11 |
| 246 | 1.55 | 0.02 | 208 | 1.59 | 0.12 |
| 244 | 1.55 | 0.03 | 207 | 1.60 | 0.14 |
| 242 | 1.56 | 0.03 | 206 | 1.60 | 0.16 |
| 240 | 1.56 | 0.04 | 205 | 1.60 | 0.18 |
| 238 | 1.56 | 0.05 | 204 | 1.60 | 0.21 |
| 236 | 1.56 | 0.06 | 203 | 1.59 | 0.24 |
| 234 | 1.56 | 0.06 | 202 | 1.57 | 0.26 |
| 233 | 1.56 | 0.07 | 201 | 1.55 | 0.28 |
| 232 | 1.56 | 0.07 | 200 | 1.52 | 0.30 |

**Figure S2: Colour map of the average of the square of the light intensity inside a 250 nm DOP droplet as a function of *n* and *k*.** The local light intensities at every point inside the droplet are squared (two-photon process) and then averaged throughout the whole droplet (average of the square of the internal light intensity). The (squared) intensity of the incident light is 1. Values above 1 mean that the average internal light intensity is enhanced and values below 1 mean that the average internal light intensity is attenuated compared with the intensity of the incident light. Enhancement can only occur for very pronounced nanofocusing (here 266 nm). Shadowing always leads to light attenuation. The calculations are for a wavelength of 207 nm, but they are very similar for other wavelengths in the region between 203 and 266 nm. The black crosses qualitatively indicate the situation for the measurements shown in the left column of Figure 4 and Figure 1c and d.

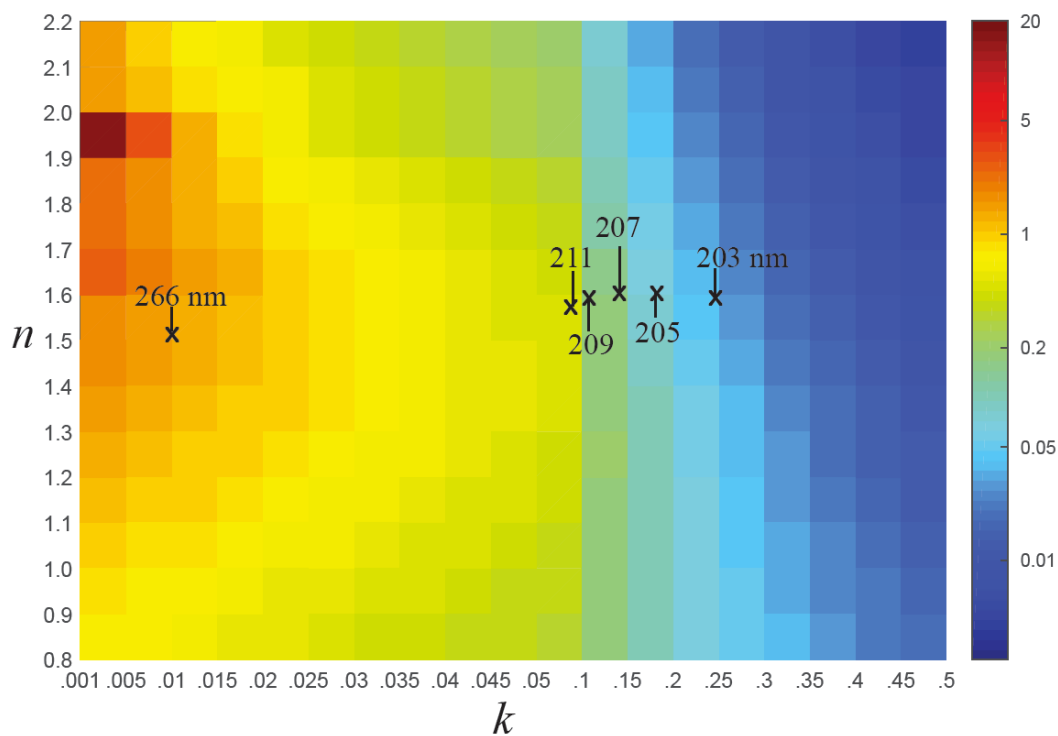

**Figure S3: Histograms of the 2D velocity distribution of the photoemission from 100 nm $H_2O$ droplets after ionization with VUV radiation of 103.3 nm.** The velocity is given in mass scaled units, i. e. $\sqrt{E_{kin}} = v_{xy}\sqrt{m_e/2}$ where $E_{kin}$ is the 2D kinetic energy as indicated in the photoelectron image (Figure 5a), $m_e$ is the mass of the electron and $v_{xy} = \sqrt{v_x^2 + v_y^2}$ is the projection of the electron velocity onto the detector plane. (a) For the experimental photoelectron image in Figure 5a. (b) For the calculated photoelectron image in Figure 5b.

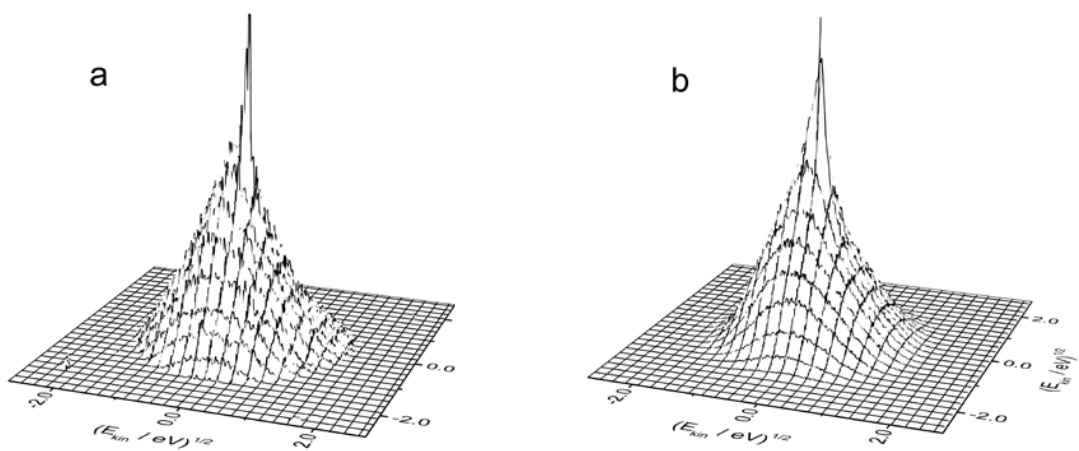

## B. Description of electron transport model

The calculation of the 2D velocity distribution used an extension of the procedure outlined in ref. [Goldmann *et al.*, J. Chem. Phys., **142**, 224304, (2015)] to account for multiple scattering processes with different cross sections, energy loss functions and anisotropies. As a starting point for the refinement we employed the model proposed by Michaud and Sanche for amorphous ice [Michaud *et al.*, Radiat. Res., **159**, 3, (2003)], which we extended to lower electron kinetic energies. The nascent kinetic energy distribution of the photoelectrons was constructed so that the photoelectron spectrum at 60 eV calculated for large droplets matches the experimental data for liquid water of ref. [Winter *et al.*, J. Phys. Chem. A, **108**, 2625, (2004)] using an escape barrier of 1eV as suggested in ref. [Michaud *et al.*, Radiat. Res., **159**, 3, (2003)]. The intensity distribution we observe experimentally in the vicinity of $v_{xy} = 0$ (Figures 5 and S3) indicates a preference for inelastic forward scattering by the escape barrier for electrons that do not match the classical escape condition for a sudden barrier (instead of elastic back scattering by specular reflection as in [Goldmann *et al.*, J. Chem. Phys., **142**, 224304, (2015)]). For the current simulation we only considered the *isotropic* component of the elastic scattering cross section. The anisotropic component is dominated by forward scattering, which does not affect the distribution of photoelectrons [Michaud *et al.*, Radiat. Res., **159**, 3, (2003)]. By fitting the experimental VMI results we refined the lowest ionization energy ($IE_0$) with its corresponding band width (half width at half maximum, $HWHM_0$), the isotropic elastic ($\sigma_e$), and the total inelastic scattering cross section ($\sigma_i$). The energy dependence of the cross sections is described be double logarithmic interpolation of the tabulated energies quoted by Michaud and Sanche [Michaud *et al.*, Radiat. Res., **159**, 3, (2003)] with a single point added at 1.0 eV and retaining the relative contributions of the different inelastic scattering processes. We followed a two-step fitting procedure: In a first step the overall 2D kinetic energy distribution (i.e. the VMI data integrated over the angular coordinate) was fitted, which allowed us to determine $IE_0$ and $HWHM_0$ while narrowing down the range of possible values for the cross sections. The latter were then refined in the second step to fit the full 2D VMI. As argued by Michaud and Sanche electron transport in liquid water is expected to be similar to that in amorphous ice, with the possible exception of the scattering off translational and librational phonons. As varying the corresponding cross sections above 1.7 eV did not significantly improve the results of the fit they were kept frozen. The final results are shown Figure S3b and Figure 5b. for $IE_0$=10.0eV, $HWHM_0$=1.2eV, $\sigma_e$=0.6Å$^2$, $\sigma_i$=1.1Å$^2$. Including the shift by the escape barrier [Elles *et al.*, J. Chem. Phys, **125**, 044515, (2006)] $IE_0$ agrees well with the lowest vertical ionization energy of liquid water. The band width we obtain (with negligible contributions from the apparatus function) is consistent with the broad band observed in the liquid microjet photoelectron spectrum of water at low excitation energies [Yamamoto *et al.*, J. Phys. Chem. A, **120**, 1153, (2016)]. Using these parameters we obtained the MFPs and electron attenuation lengths (EALs) given in Table 1. Note that EALs depend on the energy resolution $\Delta E$. We define it as "the distance at which the flux of electrons maintaining the initial kinetic energy **within $\Delta E$** diminishes by a factor of 1/e" with $\Delta E$=0.1eV.